\documentclass[review]{elsarticle}
\usepackage{amsmath}
\usepackage{amsthm}
\usepackage{amssymb}
\usepackage{titlesec}
\usepackage[thinlines,thiklines]{easybmat}

\usepackage{graphicx}
\theoremstyle{plain}\newtheorem{Df}{Definition}
\theoremstyle{plain}\newtheorem{Th}{Theorem}
\theoremstyle{definition}\newtheorem{Rm}{Remark}
\theoremstyle{definition}
\theoremstyle{plain}\newtheorem{Pp}{Proposition}
\theoremstyle{plain}
\theoremstyle{plain}\newtheorem{Lm}{Lemma}
\theoremstyle{plain}

\def\>{\ensuremath{\rangle}}
\def\<{\ensuremath{\langle}}

\def\-{\ensuremath{\textrm{-}}}

\def\fdmu{\Delta}

\def\h{\ensuremath{\mathcal{H}}}

\def\q{\bold Q}

\def\m{\ensuremath{\mathcal{M}}}

\def\s{\ensuremath{\mathcal{S}}}

\def\x{\ensuremath{\mathcal{X}}}

\def\ra{\ensuremath{\rightarrow}}

\def\Tr{\ensuremath{\mathcal{\text{Tr}}}}

\def\leqI{\ensuremath{\mathcal{SI}(\h)}}

\begin{document}
\begin{frontmatter}
\title{ Quantum  Markov chains: description of hybrid systems, decidability of  equivalence, and model checking linear-time properties }
\author[a1,a2,a4]{Lvzhou Li\corref{one}}
\author[a2,a3]{Yuan Feng}

 \cortext[one]{Corresponding author. Mobile: +8613802437672; Corresponding address: Department of
Computer Science, Sun Yat-sen University, Guangzhou 510006,
  China\\
 \indent {\it E-mail
address:}  lilvzh@mail.sysu.edu.cn (L.
Li); Yuan.Feng@uts.edu.au (Y. Feng)}

\address[a1]{Department of
Computer Science, Sun Yat-sen University, Guangzhou 510006,
  China}

\address[a2]{Center for Quantum Computation and Intelligent
Systems, FEIT, University of Technology Sydney, Sydney 2oo7, Australia}

\address[a3]{AMSS-UTS Joint Research Laboratory for Quantum Computation, Chinese Academy of Sciences, Beijing, China}

\address[a4]{The Guangdong Key Laboratory of Information Security Technology,
Sun Yat-sen University, 510006, China}

\begin{abstract}
In this paper, we study   a model of quantum Markov chains  that is a quantum analogue of  Markov chains and  is obtained by replacing  probabilities in  transition matrices with quantum operations.  We show that this model is very suited to  describe  hybrid systems that consist of a quantum component and a classical one, although it has the same expressive power as another quantum  Markov model proposed in the literature.
 Indeed, hybrid systems are often encountered in quantum information processing; for example, both quantum programs and quantum protocols can be regarded as hybrid systems. Thus, we further propose a model called hybrid quantum automata (HQA) that can be used to describe these hybrid systems that receive inputs (actions) from the outer world. We show the language equivalence problem  of HQA is decidable in polynomial time. Furthermore, we apply this result to the trace equivalence problem of  quantum  Markov chains, and thus it is also decidable in polynomial time. Finally, we discuss model checking linear-time properties of quantum  Markov chains, and show the quantitative analysis of regular safety properties can be addressed successfully.

\end{abstract}
\begin{keyword}
 Quantum  Markov chains, hybrid systems, quantum automata,  equivalence, model checking, linear-time property
\end{keyword}

\end{frontmatter}
\section{Introduction}
 As we know, Markov chains as a mathematical model for stochastic systems  play a fundamental role in computer science and even in the whole field of information science.  A Markov chain is usually  represented by a pair $( P, \pi_0)$ where $\pi_0$ is a vector standing for the initial state of a stochastic system, and $P$ is a stochastic matrix \footnote{In this paper, a  matrix is said to be a stochastic matrix if each column of it is a probabilistic distribution.} characterizing the evolution of the system. Over the past two decades, quantum computing and quantum information have attracted considerable attention from the academic community. Then it is natural to study the quantum analogue of Markov chains. Actually, the terminology ``quantum Markov chains'' have appeared many times in the literature \cite{AAKV01, AK92,  Gud08, FYY12,LP11, Ying11}, although it does not mean exactly the same thing in different references. A usual approach to defining quantum Markov chains is to view  a quantum Markov chain as a pair $( \mathcal{E}, \rho_0)$ where $\rho_0$, a density operator, denotes an initial state of a quantum system, and $\mathcal{E}$ is a trace-preserving quantum operation that characterizes the dynamics of the quantum system. This resembles very closely a classical Markov chain  represented by a pair $( P, \pi_0)$. Indeed, in the textbook \cite{NC00}, when quantum operations were introduced, they were viewed as a quantum analogue of Markov processes.  In \cite{LP11, Ying11}, a quantum Markov chain means the same thing as mentioned here, while it mainly means a quantum walk in \cite{AAKV01}.

In this paper, we  focus on the quantum Markov model reported in \cite{Gud08, FYY12} which is greatly different from the  one mentioned above but will be shown to be very suited to describe hybrid systems that consist of a quantum component and a classical one.  Such a quantum Markov chain can be roughly represented by a  pair $(M, \mu_0)$ where $M$ is a transition matrix resembling $P$ in a classical Markov chain but replacing each transition probability with a quantum operation and satisfying the condition that the sum of  each column of $M$ is a trace-preserving quantum operation. $\mu_0$, standing for the initial state of the model, is a vector with each entry being a density operator up  to a factor. This model looks very strange at first glance, but it has the same expressive power  as the conventional one given by $( \mathcal{E}, \rho_0)$. Specially, we will show that this model is very suited to describe hybrid systems that consists of a quantum component and a classical one. Indeed,  hybrid systems are often encountered in quantum computing and quantum information, varying from quantum Turing machines \cite{Wat03} and quantum finite automata \cite{LF12,QLM12} to quantum programs \cite{Sel04} and quantum protocols such as BB84. Quantum engineering systems developed  in the  future will most probably have a classical human-interactive interface and a quantum processor, and thus they will be hybrid models. Therefore, it is worth developing a theory for describing and verifying  hybrid systems.

In order to describe  hybrid systems that receive inputs or actions from the outer world, we propose the notion of hybrid quantum automata (HQA) that generalize   semi-quantum finite automata or other models studied by Ambainis
and Watrous, and Qiu etc (see e.g. \cite{AW02,BMP03, QLM15, ZQL12,ZGQ14,ZQG15}). In fact, these automata in the mentioned references  as hybrid systems have been   described in a uniform way by the authors \cite{LF12}.
When viewing HQA as language acceptors, we show their language equivalence problem is decidable  in polynomial time by transforming this problem to the equivalence problem of probabilistic automata. Furthermore, we apply this result to the trace equivalence problem of quantum Markov chains, showing the trace equivalence problem is also decidable in  polynomial time.

Finally, we consider model checking linear-time properties of  hybrid systems that are modeled by quantum Markov chains. We show that the quantitative analysis of  regular safety properties can be addressed as done for stochastic systems, by  transforming it to the reachability problem that can be addressed by determining a least solution of a  system of linear equations. For general $\omega$-regular properties, the similar technical treatments used for stochastic systems no longer take effect for our purpose, and  some new techniques need to be explored in the further study.

\section{Preliminaries}
 A
 Hilbert space is usually denoted by the symbol $\mathcal{H}$. $dim ({\cal H})$ stands for the dimension of ${\cal H}$.
  Let $\mathcal{L}({\cal H})$ be the
set of all linear operators from ${\cal H}$ to itself.
$A^*$, $A^\dagger$  and $A^\top$ denote respectively
the conjugate, the conjugate-transpose, and
the transpose  of operator $A$.   The
trace of $A$ is denoted by $\Tr(A)$.   $A\in \mathcal{L}({\cal H})$ is said to be positive, denoted by $A\geq 0$, if $\langle\psi|A|\psi\rangle\geq
     0$ for any $|\psi\rangle \in {\cal H}$. $A\geq B$ if $A-B$ is  positive. Let \begin{align*}\mathcal{P}(\mathcal{H})=\{A\in \mathcal{L}({\cal H}): A\geq 0\}.\end{align*}
Given a nonempty and countable set $S$, let
\begin{align*}Dist_{\mathcal{H}}(S)=\{\mu: S\rightarrow \mathcal{P}(\mathcal{H}): \sum_{s\in S}\Tr(\mu(s))=1\}.\end{align*}
Elements in $Dist_{\mathcal{H}}(S)$ are called {\it positive-operator valued distributions.}

 The detailed  background on quantum information can be referred to the textbook
 \cite{NC00} and lecture notes \cite{Wat}. Here we just introduce briefly some necessary
notions.
States of a quantum system are described by density operators that are positive operators having unit trace. Let
 \begin{align*}\mathcal{D}({\cal H})=\{A\in \mathcal{P}({\cal H}):  \Tr(A)=1\},\end{align*}
  which denotes the set of all density operators on Hilbert space ${\cal H}$. An element in $\mathcal{D}({\cal H})$ is generally indicated by the symbol $\rho$. A positive operator with trace less than $1$ is called a {\it partial quantum state}.

A mapping
${\cal E}$: $\mathcal{L}({\cal H})\rightarrow \mathcal{L}({\cal H})$ is
called a {\it super-operator} on ${\cal H}$. ${\cal E}$ is said to be {\it trace-preserving} if $\Tr({\cal E}(A))=\Tr(A)$  for all $A\in \mathcal{L}(\mathcal{H})$. Let $\mathcal{I}_{\cal H}$ and $\mathbf{0}_{\cal H}$ denote the identity and zero super-operators, respectively, and if $\mathcal{H}$ is clear from the context the subscript $\mathcal{H}$ is omitted. For two super-operators $\mathcal{E}$ and $ \mathcal{F}$, their
 summation, subtraction and  multiplication, denoted by $\mathcal{E}+\mathcal{F}$, $\mathcal{E}-\mathcal{F}$ and $\mathcal{E}\circ\mathcal{F}$, respectively, are defined by
\begin{align*}&(\mathcal{E}+\mathcal{F}) (A)=\mathcal{E}(A)+\mathcal{F}(A), \\
&(\mathcal{E}-\mathcal{F}) (A)=\mathcal{E}(A)-\mathcal{F}(A), \\ &\mathcal{E}\circ\mathcal{F}(A)=\mathcal{E}(\mathcal{F}(A))\end{align*}
for all $A\in\mathcal{L}({\cal H})$.
We always omit the symbol $\circ$ and write $\mathcal{E}\mathcal{F}$ simply for $\mathcal{E}\circ\mathcal{F}$.
The relation   $\eqsim$  between super-operators on ${\cal H}$ is defined by:  ${\cal E}\eqsim{\cal F}$ if $
\Tr({\cal E}(\rho))=\Tr({\cal F}(\rho))$ for all $\rho\in \mathcal{D}({\cal H})$.

The  evolution of a  quantum system is
characterized by  {\it completely positive super-operators (CPOs)} . Here we do not recall the original definition of ``completely positive'', but give an equivalent characterization.
A super-operator ${\cal E}$: $\mathcal{L}({\cal H})\rightarrow \mathcal{L}({\cal H})$  is  said to be {\it completely positive} if and only if  it has an {\it operator-sum
representation (also called Kraus representation)} as
\begin{align*}
{\cal E}(A)=\sum_kE_k A E_k^\dagger,
\end{align*}
where the set
$\{E_k\in \mathcal{L}(\mathcal{H})\}$ are called  {\it operation elements} of ${\cal E}$. ${\cal E}$ is trace-preserving if and only if its  operator-sum
representation satisfies the following {\it completeness condition}
\begin{align}
\sum_kE_k^\dagger E_k= I.\label{OPC}
\end{align}
If   Eq. (\ref{OPC}) is replaced by $\sum_kE_k^\dagger E_k\leq I$, then ${\cal E}$ is said to be {\it trace-nonincreasing}.
By $\mathcal{S}\mathcal{I}(\mathcal{H})$ we mean the set of all trace-nonincreasing CPOs on $\mathcal{H}$. Here the reason why we used the notation $\mathcal{S}\mathcal{I}(\mathcal{H})$ is to keep up with the one in \cite{FYY12}.
Elements in $\mathcal{S}\mathcal{I}(\mathcal{H})$ are called   {\it quantum operations} \cite{NC00}. Trace-preserving CPOs are called trace-preserving quantum operations.

Here we recall the concept of {\it selective quantum operations} \cite{Wat03} that are actually trace-preserving quantum operations equipped with a physical interpretation. A selective quantum operation
 is  a mapping  that takes as input $\rho\in \mathcal{D}({\cal H})$ and outputs a probability over pairs of the form $(\tau, \rho_{\tau})$, where $\tau\in \Delta$ and $\rho_\tau\in \mathcal{D}({\cal H})$. We refer to $\tau$ as the classical output of the operation; this may be the result of some measurement performed on $\rho$, but this is not the most general situation. A  selective quantum operation ${\cal E}$ is described by a set of operation elements
$$\{E_{\tau,k}:\tau\in\Delta, k\in K_\tau\}$$ which satisfy the completeness condition
$
\sum_{\tau\in\Delta}\sum_{k\in K_\tau}E_{\tau,k}^\dagger E_{\tau,k}=I.
$
For each $\tau\in \Delta$, we defined  $\Phi_\tau$ as follows:
$$\Phi_\tau(\rho)=\sum_{k\in K_\tau}E_{\tau,k}\rho E_{\tau,k}^\dagger.$$
Then   $\mathcal{E}$  is represented by  $\mathcal{E}=\{\Phi_\tau: \tau\in\Delta\}.$
For $\tau\in \Delta$, let $p_\tau=\Tr(\Phi_\tau(\rho))$ and $\rho_\tau=\Phi_\tau(\rho)/p_\tau$ (in case $p_\tau=0$, $\rho_\tau$ is undefined).
Now, on input $\rho$, the output of  $\mathcal{E}$ is defined to be $(\tau, \rho_\tau)$ with probability $p_\tau$.

 In the following, we recall a useful linear mapping $vec$ from \cite{Wat} which
maps a matrix $A\in \mathbb{C}^{n\times n}$ to an $n^2$-dimensional column vector, defined as follows:
\begin{eqnarray*}vec(A)((i-1)n+j)=A(i,j).\end{eqnarray*}
 In other words, $vec(A)$ is the vector obtained by taking the rows of $A$, transposing them to form
column vectors, and stacking those column vectors on top of one another to form a single vector. For example, we have
\begin{eqnarray*}A=\left(
                    \begin{array}{cc}
                      a & b \\
                      c & d \\
                    \end{array}
                  \right)~
                 \text{and}~ vec(A)=\left(
                    \begin{array}{cc}
                      a\\
                      b \\
                      c \\
                       d
                    \end{array}
                  \right).
\end{eqnarray*}
If we let $|i\rangle$ be an $n$-dimensional column vector with the $i$th entry being 1 and else 0's, then $\{|i\rangle\langle j|: i,j=1,\cdots, n\}$ form a basis of  $\mathbb{C}^{n\times n}$. Therefore, the mapping $vec$ can also be defined as $vec(|i\rangle\langle j|)=|i\rangle|j\rangle.$

 Let $A, B, C$ be $n\times n$ matrices.  Then we have
 \begin{eqnarray}
 vec(AXB)&=&(A\otimes B^{\top})vec(X),\label{P1}\\
  \text{Tr}(AB)&=&vec(A^{\top})^{\top}vec(B). \label{P2}
   \end{eqnarray}

Throughout this paper, we use $|S|$ to denote the cardinality of  the set $S$.

\section{Quantum  Markov chains and hybrid systems}
As mentioned in Introduction,  it is natural  to propose the quantum analogue of Markov chains, while
Markov chains have been shown to  play a fundamental role in information science and  quantum   information processing has attracted more and more attention from the academic community.  In the literature, there are several notions of  quantum  Markov chains defined from different perspectives, but in this paper we focus mainly on the one  reported in \cite{ Gud08, FYY12}. In the following, we show that the quantum Markov model given in \cite{ Gud08, FYY12} is very suited to  describe hybrid systems, although it  has the same expressive power as the conventional one given in \cite{LP11, Ying11}. Before that, we recall some necessary definitions below.

One natural viewpoint is to regard  a quantum  Markov chain as a pair of a trace-preserving quantum operation and a density operator. Here we keep up with the notations used in \cite{Ying11}.

\begin{Df}A quantum  Markov chain (qMC)  is a triple $\mathcal{M}=(\mathcal{H},\mathcal{E},\rho_0)$ where:
 \begin{itemize}
   \item $\mathcal{H}$ is a Hilbert space;
   \item $\mathcal{E}$ is a trace-preserving quantum operation over $\mathcal{H}$;
   \item $\rho_0$ is the initial density operator over $\mathcal{H}$.
 \end{itemize}
 $\mathcal{M}$ is said to be finite if  $dim(\mathcal{H})$ is finite.
\end{Df}
The physical interpretation of qMC $\mathcal{M}$ is: a quantum system  with Hilbert space $\mathcal{H}$ starts in the initial state $\rho_0$, and at each step the state evolves according to $\mathcal{E}$.
Usually, we do not give explicitly the underlying Hilbert space $\mathcal{H}$, and thus a qMC is simply denoted by the pair $(\mathcal{E},\rho_0)$. The state at the $n$th step is  denoted by $\rho_n$. Then we have \begin{align}\rho_{n}=\mathcal{E}^{n}(\rho_0)\end{align} where $\mathcal{E}^{n}$ is inductively defined by: (i) $\mathcal{E}^{0}=\mathcal{I}$  and (ii) $\mathcal{E}^{n}=\mathcal{E}\mathcal{E}^{n-1}$ for $n=1,2,\cdots$.

 Another quantum analogue of  Markov chains was reported in \cite{Gud08,FYY12}, and we call it {\it hybrid quantum Markov chain (hqMC)}. The reason why we called them ``hybrid'' will get clear soon. The definition is  as follows.
\begin{Df}
An hqMC is represented by a tuple $\mathcal{M}=(\mathcal{H}, \mathcal{S}, M, \mu_0)$ where:
 \begin{itemize}
   \item $\mathcal{H}$ is a Hilbert space;
   \item $S$ is a nonempty and countable set of states;
   \item $M:S\times S\rightarrow \mathcal{SI}(\mathcal{H})$ such that $\sum_{t\in S} M(t,s)$ is a trace-preserving quantum operation for each $s\in S$.
   \item $\mu_0\in Dist_{\mathcal{H}}(S)$ denotes the initial distribution.
 \end{itemize}
$\mathcal{M}$  is said to be finite, if $S$ and  $dim(\mathcal{H})$ are finite.
\end{Df}

\begin{figure}[htbp]\centering
\includegraphics[width=45mm]{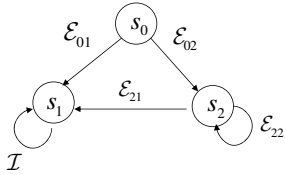}
\caption{ Transition graph of an hqMC. }\label{Tgraph}
\end{figure}

Visually, hqMC $\mathcal{M}$ can be represented by a transition graph (a digraph) where states from $S$ act as vertexes and there is an edge from $s$ to $t$ with label $M(t,s)$ if  and only if $M(t,s)\neq \mathbf{0}$. For example,  Fig. \ref{Tgraph} represents an hqMC whose state set is $S=\{s_0,s_1, s_2\}$ and whose transition function $M$ is given by $M(s_1, s_0)=\mathcal{E}_{01}, M(s_2, s_0)=\mathcal{E}_{02}, M(s_2, s_2)=\mathcal{E}_{22}, M(s_1, s_2)=\mathcal{E}_{21}, M(s_1, s_1)=\mathcal{I}$ where all given quantum operations are nonzero.

In the sequel, we  identify the transition function
$M$ with a $|S|\times |S|$ matrix  of which each entry is a quantum operation and the sum of each column is a trace-preserving quantum operation. $M(t,s)$ denotes the entry in the $t$th row and the $s$th column.   Similarly, $\mu_0$ is  viewed as a $|S|$-dimensional column vector with each entry being a positive operator and their sum being a density operator.  $\mu_0(s)$ denotes the $s$th entry. A rough  interpretation of hqMC $\mathcal{M}$ is: the system first starts in $\mu_0$ and then at each step evolves according to $M$. At the $n$th step, the state is denoted by \begin{align*}\mu_{n}= M^{n}\mu_0\end{align*}
where $M^{n}$ is inductively defined by: i) $M^{0}$ is a diagonal matrix with the diagonal entries being $\mathcal{I}$ and other entries being $\mathbf{0}$, and ii) $M^{n}=MM^{n-1}$. Note that  the multiplication of an entry in $M$ and an entry in $\mu_{n}$ is in the sense of performing a quantum operation on a partial quantum state. For example, for $\mu_{1}=M\mu_0$ we have $\mu_{1}(s)=\sum_{t\in S} M(s,t)(\mu_0(t))$ for $s\in S$ where quantum operation $M(s,t)$ is performed on the state $\mu_0(t)$.

\begin{figure}[htbp]\centering
\includegraphics[width=70mm]{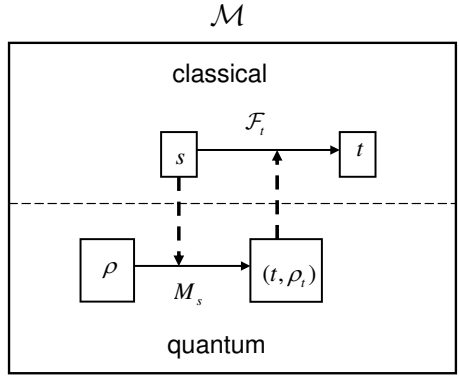}
\caption{An hqMC describing the behavior of a hybrid system.}\label{Fig-hqMC}
\end{figure}

In the following, we show that  the model of hqMC $\mathcal{M}=(\mathcal{H}, \mathcal{S}, M, \mu_0)$ is very suited to describe    hybrid systems that are dynamic systems consisting of  two interactive components: a quantum one and a classical one, although this might not be clearly noticed when the model was proposed at the beginning. As shown in Fig. \ref{Fig-hqMC},  there is  a hybrid system consisting of a quantum component whose state space is $\mathcal{H}$ and a classical component whose state set is $S$.  The behavior of this system is exactly described by $\mathcal{M}$.
More specifically, at each step the hybrid system evolves as follows.
\begin{itemize}
  \item [(i)] Firstly, depending on the current classical state $s$,  quantum state $\rho$ evolves according to $M_s$. $M_s$ denotes the selective quantum operation described by the $s$th column of $M$, that is, $M_s=\{M(t,s): t\in S\}$. Thus, on input $\rho$, the output is $(t, M(t,s)(\rho)/p_t)$ with probability $p_t=\Tr(M(t,s)(\rho))$.
  \item [(ii)] Secondly, the classical state $s$ evolves into state $t$, where $t$ is the output of the above quantum evolution. Equivalently, a transition function $\mathcal{F}_t: S\rightarrow S$  changes each  state to $t$. If $S$ is finite, then the classical component can be viewed as a DFA whose state set and input alphabet are both $S$ and whose transition function maps the current state to  the state  indicated by the current input symbol.
\end{itemize}
The initial state of the hybrid system is $(s, \mu_0(s)/p_s)$ with probability $p_s=\Tr(\mu_0(s))$ where $s\in S$ denotes the classical state.

As shown above, the hqMC model is suited to describe hybrid systems. Indeed, hybrid systems are often encountered in quantum information processing. For example, quantum programs can be regarded as hybrid systems, since  as stated by Selinger \cite{Sel04}, quantum programs can be described by ``quantum data with classical control flows''. Quantum data are represented by states of the quantum component and classical control flows are state evolutions of the classical component. Also note that the quantum Turing machines defined in \cite{Wat03} and  quantum finite automata studied in \cite{LF12} are all hybrid systems.
In addition, as shown in \cite{FYY12}, quantum cryptographic protocols such as BB84 protocol can be described by hqMC.  We think that hybrid systems will be encountered more often as the study of quantum information goes ahead. In fact, since what we can observe are  classical,  the quantum engineering systems developed in the  future
will most probably have a classical human-interactive interface and a quantum processor, and thus they will be hybrid systems.

On should distinguish   ``hybrid systems'' in this paper from those in \cite{Hen96, Ras05}. Hybrid systems in \cite{Hen96, Ras05}  are digital real-time systems embedded in analog environments. Those
systems combine discrete and continuous dynamics. There have been a long list of publications devoted to the verification of these hybrid systems.
Hybrid systems in our paper are such systems that combine classical discrete  dynamics and quantum discrete  dynamics (it is also possible to consider quantum continuous dynamics), and in the sequel, when mentioning ``hybrid systems'' we always adopt this meaning. As mentioned above, hybrid systems often present in quantum information processing. Therefore, it could  be meaningful and interesting to develop a theory of describing and verifying   these systems; Feng et al' s work \cite {FYY12} can be seen as a first step toward this direction.

In the following, we clarify the relationship between the two models of quantum  Markov chains presented in this section.
First, a qMC is obviously a special hqMC, since when there is only one classical state in an hqMC, it reduces to a qMC.
On the other hand,  we will show that each hqMC can also be simulated by a qMC. This is formally expressed in the following theorem.

\begin{Th}  Given an hqMC $\mathcal{M}=(\mathcal{H}, S, M, \mu_0)$, there exists a qMC $\mathcal{M}'=(\mathcal{H}',\mathcal{E},\rho_0)$ such that the states   $\rho_n$ and $\mu_n$ of $\mathcal{M}'$ and $\mathcal{M}$ at the $n$th step, respectively, satisfy:
\begin{align*}\rho_{n}=\sum_{s\in S}|s\rangle\langle s|\otimes \mu_{n}(s)\end{align*} for $n=0,1,\cdots.$ \label{Th:1}
\end{Th}
{\noindent\it Proof.} The qMC $\mathcal{M}'=(\mathcal{H}',\mathcal{E},\rho_0)$ is constructed  as follows:
\begin{itemize}
  \item ${\cal H}'=\mathcal{H}_S\otimes \mathcal{H}$ where $\mathcal{H}_S=span\{|s\rangle: s\in S\}$;
  \item $\rho_0=\sum_{s\in S}|s\rangle\langle s|\otimes \mu_0(s)$;
  \item $\mathcal{E}$ is a trace-preserving quantum operation on ${\cal H}'$ which is constructed to simulate the interactive actions between the quantum component and the classical one in Fig. \ref{Fig-hqMC}.
\end{itemize}
More specifically, $\mathcal{E}$ is described by the set of operation elements \begin{align}\{|t\rangle\langle s|\otimes M_{ts}^k: s,t\in S, k\in K_{ts}\}\end{align} where
  for each pair $t,s\in S$, $\{M_{ts}^k: k\in K_{ts}\}$ are operation elements of the quantum operation $M(t,s)$.

Then $\mathcal{E}$  is trace-preserving since we have
\begin{align*}
&\sum_{s,t\in S}\sum_{k\in K_{ts}} (|t\rangle\langle s|\otimes M_{ts}^k)^\dagger (|t\rangle\langle s|\otimes M_{ts}^k)\\
=&\sum_{s,t\in S}\left(|s\rangle \langle s| \otimes \sum_{k\in K_{ts}} (M_{ts}^k)^\dagger M_{ts}^k\right)\\
=&\sum_{s\in S}\left(|s\rangle \langle s|\otimes \sum_{t\in S}\sum_{k\in K_{ts}} (M_{ts}^k)^\dagger M_{ts}^k\right)\\
\stackrel{(a)}{=}&\sum_{s\in S}|s\rangle \langle s|\otimes I_{\mathcal{H}}=I_{\mathcal{H}_S}\otimes I_{\mathcal{H}}
\end{align*}
where $I_{\mathcal{H}_S}$ and $I_{\mathcal{H}}$ denote  identity operators on $\mathcal{H}_S$ and $\mathcal{H}$, respectively. In the above, equation (a) holds because  $\{M_{ts}^k: k\in K_{ts}, t\in S\}$ are operation elements of the selective quantum operation $M_s$.

Furthermore, for $\rho\otimes\varrho \in \mathcal{L}(\mathcal{H}_S\otimes \mathcal{H})$, by a direct calculation we have
\begin{align}\mathcal{E}(\rho\otimes\varrho)&=\sum_{s,t\in S}\langle s|\rho |s\rangle |t\rangle\langle t|\otimes M(t,s)(\varrho)\label{Eq6}\\
&=\sum_{s,t\in S}\langle s|\rho |s\rangle \mathcal{F}_{t}(|s\rangle\langle s|)\otimes M(t,s)(\varrho)
.\end{align}
From the above, it can be seen that the intuitive idea of $\mathcal{E}$ is as follows: i) first perform the measurement $\{E_s=|s\rangle\langle s|: s\in S\}$ on the classical system to observe its state; ii) if  $s$ is the result, then perform the selective quantum operation $M_s=\{M(t,s): t\in S\}$ on the quantum system; iii) if the classical output of $M_s$ is $t$, then perform $\mathcal{F}_{t}$ on the classical system changing its state to $t$. Note that $\mathcal{F}_{t}$ is a trace-preserving quantum operation for each $t\in S$,  since it has operation elements $\{ F^s_{t}=|t\rangle \langle s|, s\in S\}$ that satisfy the completeness condition.

Now by induction on $n$ we prove  $\rho_n=\sum_{s\in S}|s\rangle\langle s|\otimes \mu_{n}(s)$ for $n=0,1,\cdots$.
First when $n=0$, we have $\rho_0=\sum_{s\in S}|s\rangle\langle s|\otimes \mu_0(s)$. Suppose it holds for $n$. Then from Eq. (\ref{Eq6}) we have
\begin{align*}
\rho_{n+1}&=\mathcal{E}(\rho_{n})=\mathcal{E}\left(\sum_{t\in S}|t\rangle\langle t|\otimes \mu_{n}(t)\right)\\
&=\sum_{s,s'\in S}\sum_{t\in S}\langle s|t\rangle\langle t|s\rangle |s'\rangle\langle s'|\otimes M(s',s)(\mu_{n}(t))\\
&=\sum_{s'\in S}|s'\rangle\langle s'|\otimes \sum_{t\in S} M(s',t)(\mu_{n}(t))\\
&=\sum_{s'\in S}|s'\rangle\langle s'|\otimes \mu_{n+1}(s').
\end{align*}
Thus, we have completed the proof. \qed\\

From the above discussion, we know that qMC and hqMC have the same  expressive power. However, they are suitable for describing different systems.  While it is natural to describe a purely quantum system using the qMC model, it is convenient to describe a hybrid system using the hqMC model.

\section{Hybrid quantum automata}

Based on the hqMC model, we propose an  automaton model---hybrid quantum  automata, which generalizes the models in \cite{LF12} and  can be used to describe  hybrid systems that receive inputs (or actions) from the outer world.

First, some notations are explained below. As usual, for nonempty set
$\Sigma$, by $\Sigma^{*}$ we mean the set of all finite-length
strings over $\Sigma$. Let $\Sigma^+=\Sigma^{*}\setminus\{\epsilon\}$  where $\epsilon$ denotes the empty string. For $u\in \Sigma^{*}$,
$|u|$ denotes the length of $u$. Let $\Sigma^n=\{u\in \Sigma^{*}: |u|=n\}$ and $\Sigma^{\leq n}=\{u\in \Sigma^{*}: |u|\leq n\}$.

\begin{Df}
A hybrid quantum   automaton (HQA) is a tuple $$\mathcal{A}=(\mathcal{H}, S,  \Sigma, \mu_0, \{M_\sigma\}_{\sigma\in \Sigma})$$ where
\begin{itemize}
\item $\mathcal{H}$ is a Hilbert space;
  \item $S$ is a countable nonempty  set of states;
   \item $\Sigma$ is an  alphabet of symbols;
    \item $\mu_0\in Dist_{\mathcal{H}}(S)$ is the initial distribution;
  \item For each $\sigma\in \Sigma$, $M_\sigma:S\times S\rightarrow \mathcal{SI}(\mathcal{H})$  such that $\sum_{t\in S} M_\sigma(t, s)$ is a trace-preserving quantum operation for each $s\in S$;
  \end{itemize}
$\mathcal{A}$ is said to be finite, if $S$, $\Sigma$ and   $dim(\mathcal{H})$ are finite.
\end{Df}
The behavior of $\mathcal{A}$ is roughly as: $\mathcal{A}$ starts in $\mu_0$, and at each step, it scans the current input symbol $\sigma\in\Sigma$, and then updates its state according to $M_\sigma$. In this paper, we regard HQA $\mathcal{A}$ as a language acceptor, that is,  for each input  $w\in\Sigma^*$, $\mathcal{A}$  observes  its final state after scanning all  input symbols and  accepts if the final state satisfies some given property. Generally, the accepting behavior is probabilistic, because of the inherent probabilism of quantum mechanics.

Here, we have two  basic approaches to defining the automaton's accepting fashions. One is based on  classical states: $\mathcal{A}$ accepts its input $w\in \Sigma^*$, if its classical state after scanning the whole input belongs to a subset $F\subseteq S$. In this case,   the model is represented by a tuple $\mathcal{A}=(\mathcal{H}, S,  \Sigma, \mu_0, \{M_\sigma\}_{\sigma\in \Sigma},F)$, $\mathcal{A}$ is said to {\it accept with classical fashion}, and we call it a C-HQA for short.
Then  C-HQA $\mathcal{A}$ defines a function $P_{\mathcal{A}}: \Sigma^*\rightarrow [0,1]$ as
\begin{align*} P_{\mathcal{A}}(w)=\sum_{s\in F}\Tr((M_w \mu_0)(s))\end{align*}
where $M_{\sigma_1 \sigma_1\cdots \sigma_n}=M_{\sigma_n}\cdots M_{\sigma_2}M_{\sigma_1}$ and $M_\epsilon$ is a diagonal matrix with the diagonal entries being $\mathcal{I}$ and others being $\mathbf{0}$.  $P_{\mathcal{A}}(w)$ denotes the probability that $\mathcal{A}$ accepts $w$.

Also, we can define that HQA $\mathcal{A}$ accepts its input if its final quantum state  belongs to a subspace of $\mathcal{H}$, say $\mathcal{H}_{acc}$. Let $P_{acc}$ be the projector onto $\mathcal{H}_{acc}$. In this case, the model is given by $\mathcal{A}=(\mathcal{H}, S,  \Sigma, \mu_0, \{M_\sigma\}_{\sigma\in \Sigma}, P_{acc})$, $\mathcal{A}$ is said to {\it accept with quantum fashion}, and we call it a Q-HQA for short.  The probability that  $\mathcal{A}$ accepts its input $w\in \Sigma^*$ is given by \begin{align*} P_{\mathcal{A}}(w)=\sum_{s\in S}\Tr(P_{acc}(M_w \mu_0)(s)).\end{align*}

Based on the above two basic accepting fashions, HQA  can generally have a mixed accepting fashion. In this case, the model is called M-HQA for short and is represented  by $\mathcal{A}=(\mathcal{H}, S,  \Sigma, \mu_0, \{M_\sigma\}_{\sigma\in \Sigma}, F, P_{acc})$. The accepting probability on input $w\in \Sigma^*$ is give by
\begin{align*} P_{\mathcal{A}}(w)=\sum_{s\in F}\Tr(P_{acc}(M_w \mu_0)(s)).\end{align*}

\begin{Rm} (i) It is obvious that a probabilistic automaton is a degenerate  C-HQA in which $M(s,t)=p_{s,t}\mathcal{I}$ for all  $s,t\in S$ with $(p_{s,t})_{s,t\in S}$ being a stochastic matrix, and $\mu_0(s)=p_s\rho$ for some density operator $\rho$ and $s\in S$ with $(p_s)_{s\in S}$ being a  probabilistic distribution. (ii) Note that  we characterized three models of quantum finite automata  in the framework of hybrid systems in \cite{LF12};   all of them can be regarded as a concrete implementation of the HQA model defined in this paper.  For example, CL-1QFA \cite{BMP03} and 1QCFA \cite{ZQL12} are instances of C-HQA, and 1QFAC  \cite{QLM15} are instances of Q-HQA. Thus, the HQA   given by us is a  generalized model.
\end{Rm}

Associated with the qMC model, there is another quantum automaton model that was studied in \cite{LQ12,Hir08b}.
\begin{Df} A quantum automaton (QA) is a tuple $\mathcal{A}=(\mathcal{H}, \Sigma, \rho_0, \{\mathcal{E}_\sigma\}_{\sigma \in \Sigma}, P_{acc})$ where $\mathcal{H}$ is a Hilbert space, $\Sigma$ is an alphabet, $\rho_0\in \mathcal{D}(\mathcal{H})$ is the initial state, $\mathcal{E}_\sigma$ is a trace-preserving quantum operation for each $\sigma\in \Sigma$, $P_{acc}$ denotes a projector onto a subspace of $\mathcal{H}$ (called an accepting subspace). $\mathcal{A}$ is said to be finite, if $\Sigma$ and  $dim(\mathcal{H})$ are finite.
For each input $w=\sigma_1\cdots \sigma_k\in \Sigma^*$, the accepting probability is given by $P_\mathcal{A}(w)=\Tr(P_{acc}\mathcal{E}_w(\rho_0))$ where $\mathcal{E}_w=\mathcal{E}_{\sigma_k}\cdots \mathcal{E}_{\sigma_1}$.
\end{Df}

Implied by Theorem \ref{Th:1}, we have the following result.
\begin{Lm} For each HQA $\mathcal{A}$ over alphabet $\Sigma$, there is a QA $\mathcal{A}'$ such that $P_\mathcal{A}(w)=P_{\mathcal{A}'}(w)$ for all $w\in \Sigma^*$. \label{Lm:1}
\end{Lm}
{\noindent \it Proof.} The idea is similar to the procedure of simulating hqMC by qMC, and we sketch it as follows. Given an HQA $\mathcal{A}=(\mathcal{H}, S,  \Sigma, \mu_0,\{M_\sigma\}_{\sigma\in \Sigma})$, we construct a QA $\mathcal{A}'=(\mathcal{H}', \Sigma, \rho_0, \{\mathcal{E}_\sigma\}_{\sigma \in \Sigma}, P_{acc}')$ where $\mathcal{H}'= \mathcal{H}_S\otimes \mathcal{H}$, $\rho_0=\Sigma_{s\in S}|s\rangle\langle s|\otimes \mu_0(s)$, and for each $\sigma\in \Sigma$, $\mathcal{E}_\sigma$ is constructed from $M_\sigma$ as done in Theorem \ref{Th:1}. Let $\rho_w=\mathcal{E}_w(\rho_0)$ and $\mu_w=M_w\mu_0$ with $w\in\Sigma^*$. Then  as shown in Theorem \ref{Th:1} we have $\rho_w=\sum_{s\in S}|s\rangle \langle s|\otimes \mu_w(s)$. The last step is to construct $P_{acc}'$, which is dependent on the accepting fashion of $\mathcal{A}$: \begin{itemize}
\item [(a)]If $\mathcal{A}$ is a C-HQA and assume that its classical accepting set is $F\subseteq S$, then we let $P_{acc}'=\sum_{s\in F}|s\rangle \langle s| \otimes I_{\mathcal{H}}$.
\item [(b)] If $\mathcal{A}$ is a Q-HQA with projector  is $P_{acc}$, then we let $P_{acc}'=I_{\mathcal{H}_S} \otimes P_{acc}$.
 \item [(c)] If $\mathcal{A}$ is an M-HQA, then let   $P_{acc}'=\sum_{s\in F}|s\rangle \langle s| \otimes P_{acc}$.
                                           \end{itemize}
In any case, it is easy to verify that $P_\mathcal{A}(w)=P_{\mathcal{A}'}(w)$ for all $w\in \Sigma^*$. \qed\\
\begin{Rm}
From the above result it follows that HQA do not surpass QA in the sense of language recognition power.  Note that  it has been shown that finite QA  recognize with bounded error exactly the family of regular language\cite{LQ12}.
\end{Rm}

In the following we introduce another model that was called bilinear machine in \cite{LQ08a}.
\begin{Df} A bilinear machine (BLM) is a tuple $\mathcal{A}=(n, \Sigma, \{M_\sigma\}_{\sigma\in\Sigma}, \pi,\eta)$ where $n\in\mathbb{N}$ is called the state number of $\mathcal{A}$, $\Sigma$ is a finite alphabet, $M_\sigma\in \mathbb{C}^{n\times n}$ is a transition matrix for each $\sigma\in\Sigma$,  $\pi\in \mathbb{C}^{n}$ is a column vector, and $\eta\in \mathbb{C}^{n}$ is a row vector. Automaton $\mathcal{A}$ assigns each $w=\sigma_1\cdots \sigma_k\in \Sigma^*$ a weight $P_{\mathcal{A}}(w)$ as $P_{\mathcal{A}}(w)=\eta M_{\sigma_k}\cdots M_{\sigma_1}\pi$.   $\mathcal{A}$ is a probabilistic automaton if it is further required that each $M_\sigma$ is a stochastic matrix, $\pi$ is a probabilistic distribution, and $\eta$ has entries being $1$ or $0$.
\end{Df}

Every finite QA can be simulated by a BLM, which is stated formally as follows.
\begin{Lm} For each finite QA $\mathcal{A}$ over alphabet $\Sigma$, there is a  BLM $\mathcal{A}'$ such that $P_{\mathcal{A}'}(w)=P_\mathcal{A}(w)$ for all $w\in \Sigma^*$. \label{Lm:2}
\end{Lm}

{\noindent\it Proof.} Let finite QA $\mathcal{A}=(\mathcal{H}, \Sigma, \rho_0, \{\mathcal{E}_\sigma\}_{\sigma \in \Sigma}, P_{acc})$. For each $\sigma\in \Sigma$, suppose that ${\cal E}_{\sigma}(\rho)=\sum_k E^{\sigma}_k\rho{E^{\sigma}_k}^\dagger$, and denote \begin{equation*}A_{\sigma}=\sum_kE^{\sigma}_k\otimes {E^{\sigma}_k}^*.\end{equation*}
  Then by Eq. (\ref{P1}), we have
  \begin{eqnarray*}
  vec({\cal E}_{\sigma_1}(\rho))&=&A_{\sigma_1}vec(\rho),\\
  vec({\cal E}_{\sigma_2}{\cal E}_{\sigma_1}(\rho))&=&A_{\sigma_2}A_{\sigma_1}vec(\rho).
\end{eqnarray*}
As a result, the probability of ${\cal A}$ accepting $w=\sigma_1\cdots, \sigma_k\in \Sigma^*$ can be rewritten in the following:
 \begin{align*}
 P_{\mathcal{A}}(w)&=\Tr(P_{acc}\mathcal{E}_{\sigma_k}\cdots \mathcal{E}_{\sigma_1}(\rho_0))\\
 &=vec(P_{acc})^{\top} vec(\mathcal{E}_{\sigma_k}\cdots \mathcal{E}_{\sigma_1}(\rho_0))\\
 &=vec(P_{acc})^{\top} A_{\sigma_k}\cdots A_{\sigma_2}A_{\sigma_1}vec(\rho_0)
\end{align*}
where the second equality follows from Eq. (\ref{P2}).
Therefore, we construct BLM  $\mathcal{A}'=(n, \Sigma, \{M_\sigma\}_{\sigma\in\Sigma}, \alpha,\eta)$ with $n=dim(\mathcal{H})^2$, $M_\sigma=A_\sigma$ for $\sigma\in\Sigma$, $\pi=vec(\rho_0)$, and $\eta=vec(P_{acc})^{\top}$.
\qed\\

In classical automata theory, it is a fundamental problem to decide wether two probabilistic automata have the same accepting probability for each input (that is known as the equivalence problem) \cite{Paz71,Tze92,KMQ11}. This problem has some nontrivial applications; for example, \cite{CC92} applied it to verification of equivalence between processes, and \cite{MQ05,KMQ12} applied it to verification of equivalence between probabilistic programs. Taking account into that the model of HQA is suited to describe  hybrid systems (including quantum programs), it is meaningful to consider the equivalence problem for HQA. Formally,  the equivalence problem is as follows.
\begin{Df}Two HQA (QA, BLM) $\mathcal{A}_1$ and $\mathcal{A}_2$ over the same alphabet $\Sigma$ are $k$-equivalent, if $P_{\mathcal{A}_1}(w)=P_{\mathcal{A}_2}(w)$ for all $w\in \Sigma^{\leq k}$. Furthermore, they are said to be equivalent if $P_{\mathcal{A}_1}(w)=P_{\mathcal{A}_2}(w)$ holds for  all $w\in \Sigma^*$. \label{Df-eq}
\end{Df}

The history for  the equivalence problem of probabilistic automata is as follows. Paz \cite{Paz71} proved that two probabilistic automata are equivalent if and only if they are $(n_1+n_2-1)$-equivalent, where $n_1$ and $n_2$ are state numbers of the two automata. Afterwards, this result was improved by Tzeng \cite{Tze92} who proposed a polynomial-time algorithm determining whether two  given probabilistic automata are equivalent or not, and the time complexity is $O(|\Sigma|(n_1+n_2)^4)$. Recently, an improved complexity  $O(|\Sigma|(n_1+n_2)^3)$ was reported  in \cite{KMQ11}.
As mentioned in  \cite{LQ08a}, all  these results   are  based on some ordinary knowledge about matrices and linear spaces rather than on any essential property of probabilistic automata; as a result, they also hold for BLM.
We summarize these results  as follows.

\begin{Lm} Two BLM  $\mathcal{A}_1$ and $\mathcal{A}_2$ over $\Sigma$ are equivalent if and only if they are $(n_1+n_2-1)$-equivalent, and there exists a $O(|\Sigma|(n_1+n_2)^3)$ time algorithm deciding whether they are equivalent or not, where $n_1$ and $n_1$ are state numbers of  $\mathcal{A}_1$ and $\mathcal{A}_2$, respectively. \label{Lm:3}
\end{Lm}
For the sake of completeness, we present an algorithm for  BLM's equivalence problem in  Algorithm 1. For algorithmic purposes we assume that all inputs consist of
complex numbers whose real and imaginary parts are rational
numbers and that each arithmetic operation on rational numbers can
be done in constant time. Let $\mathcal{Q}$ be the set of vectors that have been  added into $queue$, and
let $$\mathcal{S}=span\left\{\left(
                         \begin{array}{c}
                           M_w^1\pi_1 \\
                            M_w^2\pi_2 \\
                         \end{array}
                       \right): w\in\Sigma^*
\right\}$$ where $M_w^1=M^1_{w_k}\cdots M^1_{w_2}M^1_{w_1}$ for $w=w_k\cdots w_2w_1$ and it is similar for $M_w^2$. Then the relationship among $\mathcal{B}$, $\mathcal{Q}$ and $\mathcal{S}$ is:  $\mathcal{Q}$ can be proved to be a basis for $\mathcal{S}$, and $\mathcal{B}$ is obtained from $\mathcal{Q}$ by the Gram-Schmidt  procedure  and thus is an orthonormal basis for $\mathcal{S}$. Therefore,  $\mathcal{A}_1$ and $\mathcal{A}_2$ are equivalent (i.e.,  $\eta_1u_1=\eta_2u_2$ holds for all elements $\left(\begin{array}{c}
                                                                    u_1 \\
                                                                    u_2
                                                                  \end{array}\right)
\in \mathcal{S}$) if and only if  $\eta_1u_1=\eta_2u_2$ for all elements $\left(\begin{array}{c}
                                                                    u_1 \\
                                                                    u_2
                                                                  \end{array}\right)
\in \mathcal{Q}$. Let $n=n_1+n_2$. Then $|\mathcal{Q}|=|\mathcal{B}|$ is at most $n$ and the  procedure  $b:=u-\sum_{b_i\in \mathcal{B}}(u^*b_i) b_i$ takes at most $O(n^2)$  time. The total time complexity is thus $O(|\Sigma|n^3)$.

\begin{center}
 \fbox{\parbox{12cm}{
 {\small
\begin{quote} {\bf Input:}  ${\cal
A}_i=(n_i, \Sigma, \{M^i_\sigma\}_{\sigma\in\Sigma}, \pi_i,\eta_i)$ for $i=1,2$.\\
{\bf Output:}  ${\cal
A}_1$ and ${\cal
A}_2$ are equivalent or not.
\begin{quote}
  ${\cal B}:=\varnothing$; $queue:=\varnothing$;
 $\pi:=\left(
          \begin{array}{c}
            \pi_1 \\
            \pi_2 \\
          \end{array}
        \right)$;\\
{\bf If} $\eta_1 \pi_1\neq \eta_2 \pi_2$ {\bf
then}\begin{quote} return ``${\cal
A}_1$ and ${\cal
A}_2$ are  not equivalent'';\end{quote}
{\bf If} $||\pi||=0$ {\bf
then}\begin{quote} return ``${\cal
A}_1$ and ${\cal
A}_2$ are  equivalent'';\end{quote}
${\cal B}:=\{\frac{\pi}{||\pi||}\}$; add $\pi$ to $queue$;\\
 {\bf while} $queue\neq \varnothing$ {\bf do}\\
 {\bf begin} take $\left(
                     \begin{array}{c}
                       v_1 \\
                       v_2 \\
                     \end{array}
                   \right)
$ from $queue$;
 \begin{quote}
 for all $\sigma\in \Sigma$  {\bf do}\\
{\bf begin}
\begin{quote}
$u_1=M^1_\sigma v_1$; $u_2=M^2_\sigma v_2$;\\
{\bf if} $\eta_1u_1\neq \eta_2u_2$ {\bf then}  \begin{quote} return ``${\cal
A}_1$ and ${\cal
A}_2$ are not equivalent'';\end{quote}
$u:=\left(
          \begin{array}{c}
            u_1 \\
            u_2 \\
          \end{array}
        \right)$;\\
$b:=u-\sum_{b_i\in \mathcal{B}}(u^*b_i) b_i$;\\
{\bf if} $||b||\neq 0$ {\bf then}
\begin{quote}
add $u$ to $queue$;\\
$\mathcal{B}:=\mathcal{B}\cup\{\frac{b}{||b||}\}$;
\end{quote}
\end{quote}
{\bf end};
 \end{quote}
 {\bf end};\\return ``${\cal
A}_1$ and ${\cal
A}_2$ are  equivalent'';
\end{quote}

\end{quote}

}
 }
 } \\\vskip 2mm
 Algorithm 1: Determining whether two BLM are equivalent or not.
\end{center}

\begin{Rm} In Definition \ref{Df-eq}, if it is required that  $P_{\mathcal{A}_1}(w)=P_{\mathcal{A}_2}(w)$ holds for  all $w\in \Sigma^+$ instead of for all $w\in \Sigma^*$, then the algorithm almost keeps the same and the  complexity has no change. This case will be used in the next section for the trace equivalence problem of quantum Markov chains.
\end{Rm}

Now it follows from Lemmas \ref{Lm:1}, \ref{Lm:2} and \ref{Lm:3} that  the equivalence problem of finite HQA is decidable in polynomial time.
\begin{Th} Two finite HQA  $\mathcal{A}_i=(\mathcal{H}^{(i)}, S^{(i)}, \mu_0^{(i)}, \Sigma, \{M_\sigma^{(i)}\}_{\sigma\in \Sigma})$ $(i=1,2)$ are equivalent if and only if they are $((n_1k_1)^2+(n_2k_2)^2-1)$-equivalent where $n_i=dim(\mathcal{H}^{(i)})$ and $k_i=|S^{(i)}|$. Furthermore, there exists a polynomial-time algorithm deciding whether they are equivalent or not.\label{Th:2}
\end{Th}

In the next section, we will show that the trace equivalence problem of  quantum Markov chains can be transformed in linear time to the equivalence problem of QHA, and thus is also decidable in polynomial time.

\section{Trace equivalence of quantum Markov chains}
Feng et al \cite{FYY12} used the model of hqMC for model-checking quantum protocols where the purpose is  to check whether the classical component of a hybrid system satisfies some given property. To that end,  a labeling function was used to associate each classical state  a set of atomic propositions that are satisfied at that state.  In this paper, we called  such an hqMC  equipped  with a state labeling function   a {\it state-labeled hybrid quantum  Markov chain} (SL-hqMC, for short).  Also, we require that the hqMC is finite, although finiteness is not a necessary requirement for a general definition. The formal definition is as follows.
\begin{Df}
A  state-labeled hybrid quantum Markov chain (SL-hqMC) is a tuple $$\mathcal{M}=(\mathcal{H}, S, M, \mu_0, AP, L),$$ where
\begin{enumerate}
\item $(\mathcal{H}, S, M, \mu_0)$ is a finite hqMC;
\item $AP$ is a finite set of atomic propositions;
\item $L: S\rightarrow 2^{AP}$ is a labeling function. $L$ can be extended to finite sequence of states as $L(s_0s_1\dots s_n)=L(s_0)L(s_1)\cdots L(s_n)$.
\end{enumerate}
\end{Df}

 In the sequel, for the sake of simplicity
 we let $\Sigma=2^{AP}$, and call it a labeling set. Then $L$ assigns each state $s\in S$ a symbol $\sigma\in \Sigma$. For $\bar{s}=s_0s_1\cdots s_k\in S^+$, let
\begin{align}\rho_{\bar{s}}=\prod_{i=1}^k M(s_i,s_{i-1})(\mu_0(s_0)) \label{state}\end{align}
where  $\prod_{i=1}^n A_i=A_n\cdots A_2 A_1$. Then $\Tr(\rho_{\bar{s}})$
 gives the probability of visiting the sequence of states $s_0s_1\cdots s_k$ when $\mathcal{M}$ starts in the initial distribution $\mu_0$.
Thus, SL-hqMC $\mathcal{M}$ defines a function $P_{\mathcal{M}}:\Sigma^+\rightarrow [0,1]$  by \begin{align*}P_{\mathcal{M}}(w)=\sum_{ \bar{s}: L(\bar{s})=w}\Tr(\rho_{\bar{s}}).\end{align*}
 This gives the probability of observing $w\in \Sigma^+$ when $\mathcal{M}$ starts in the initial distribution $\mu_0$.

In the following, we consider the trace equivalence problem of  SL-hqMC.  As shown in \cite{BK08}, the issue of trace equivalence is closely related to model checking linear-time properties  of nonprobabilistic transition systems.
For  probabilistic systems,  this problem was also discussed in \cite{DHR08}. The definition of trace equivalence is as follows.

\begin{Df}
Two SL-hqMC $\mathcal{M}_1$ and $\mathcal{M}_2$ with the same labeling set $\Sigma$ are trace equivalent if $P_{\mathcal{M}_1}(w)=P_{\mathcal{M}_2}(w)$ for all $w\in \Sigma^+$.
\end{Df}
 In the following, we  transform the trace equivalence problem of SL-hqMC to the equivalence problem of finite C-HQA that is decidable in polynomial time as shown in Theorem \ref{Th:2}.
\begin{Lm}
For every SL-hqMC $\mathcal{M}$ with labeling set $\Sigma$, we can construct in linear time a finite C-HQA $\mathcal{A}$ such that $P_{\mathcal{A}}(w)=P_{\mathcal{M}}(w)$ for all $w\in \Sigma^+$. \label{Lm:4}
\end{Lm}
{\it Proof.} Let $\mathcal{M}=(\mathcal{H}, S, M, \mu_0, AP, L)$ be an SL-hqMC. Note that $\Sigma=2^{AP}$. We construct C-HQA $\mathcal{A}=(\mathcal{H}, S', \Sigma,  \mu'_0, \{M_\sigma\}_{\sigma\in \Sigma}, F)$  as follows.
\begin{itemize}
  \item $S'=S\cup\{\tau\}$;
  \item $\mu'_0(s)=\mu_0(s)$ for all $s\in S$ and $\mu'_0(\tau)$ is the zero operator in $\mathcal{P}(\mathcal{H})$;
     \item $F=S$;
  \item For each $\sigma\in \Sigma $, $M_\sigma$ is constructed as
  \begin{align*}
  M_\sigma=\hat{M}D_\sigma
  \end{align*}\end{itemize}
where $\hat{M}$ is a block matrix:
 \begin{align*}\hat{M}=\left(
           \begin{array}{cc}
             M & \mathbf{0} \\
             \mathbf{0} & \mathcal{I} \\
           \end{array}
         \right).
  \end{align*}
That is, $\hat{M}(s,t)=M(s,t)$ for $s,t\in S$, $\hat{M}(\tau,\tau)=\mathcal{I}$, and others are $\mathbf{0}$.
 $D_\sigma$ is given by
\begin{align*}D_\sigma= \left(
              \begin{array}{ccccc}
                \delta_{L(s_1),\sigma}\mathcal{I} & \mathbf{0} &\ldots & \ldots & \mathbf{0} \\
               \mathbf{ 0} & \delta_{L(s_2),\sigma}\mathcal{I} & \mathbf{0} &\ldots & \mathbf{0} \\
                \vdots & \mathbf{0} & \ddots & \mathbf{0} & \vdots \\
                \mathbf{0} & \ldots& \mathbf{0} & \delta_{L(s_n),\sigma}\mathcal{I} & \mathbf{0 }\\
               \overline{\delta}_{L(s_1),\sigma}\mathcal{I} & \overline{\delta}_{L(s_2),\sigma}\mathcal{I} & \ldots & \overline{ \delta}_{L(s_n),\sigma}\mathcal{I} & \mathcal{I} \\
              \end{array}
            \right).\end{align*}
That is, $D_\sigma(s,s)=\delta_{L(s),\sigma}\mathcal{I}$, $D_\sigma(\tau,s)=\overline{\delta}_{L(s),\sigma}\mathcal{I}$ for all $s\in S$,  $D_\sigma(\tau,\tau)=\mathcal{I}$, and others are $\mathbf{0}$.   In the above, for  $s\in S$ and $\sigma\in \Sigma$, the meaning of $\delta_{L(s),\sigma}$ is
$$\delta_{L(s),\sigma}=\left\{
  \begin{array}{ll}
    1, & \hbox{if $L(s)=\sigma$;} \\
    0, & \hbox{otherwise.}
  \end{array}
\right.$$
In addition, $\overline{\delta}_{L(s),\sigma}=1-\delta_{L(s),\sigma}$.

It is obvious that $\hat{M}$ and $D_\sigma$ satisfy the property that the sum of each column is a trace-preserving quantum operation, and  their multiplication also satisfies this property.

   By the above construction, $\mathcal{A}$ satisfies the following property.
\begin{Pp} Let $w=\sigma_0\sigma_1\cdots \sigma_k\in \Sigma^{k+1}$ and $\mu'_w=M_w \mu'_0$. Then for any $s\in S$, we have \begin{align}\mu'_w(s)=\sum_{L(s_0\cdots s_k)=w}  M(s,s_k) \prod^k_{i=1}M(s_i,s_{i-1})(\mu'_0(s_0)),\label{pp}\end{align}
where  $L(s_0\cdots s_k)=w$  stands for ``$s_0,\cdots, s_k\in S: L(s_0\cdots s_k)=w$'' and we will always adopt this succinct notation in the sequel.

\end{Pp}
{\noindent \it Proof.} We prove it by induction on $k$.
When $k=0$, for $s\in S$ we have
\begin{align*}\mu'_{\sigma_0}(s)&=(M_{\sigma_0}\mu'_0)(s)=(\hat{M}D_{\sigma_0}\mu'_0)(s)\\
&=\sum_{s_0\in S'}\hat{M}(s,s_0)(D_{\sigma_0}\mu'_0)(s_0)\\
&=\sum_{s_0\in S}M(s,s_0)(D_{\sigma_0}\mu'_0)(s_0)\\
&=\sum_{s_0\in S}\delta_{L(s_0)\sigma_0}M(s,s_0)(\mu'_0(s_0))\\
&=\sum_{L(s_0)=\sigma_0}M(s,s_0)(\mu'_0(s_0)).
\end{align*}
Suppose the result holds for $k$. Then for $s\in S$ and $\sigma_{k+1}\in\Sigma$ we have
\begin{align*}
&\mu'_{w\sigma_{k+1}}(s)=(M_{\sigma_{k+1}}\mu'_w)(s)=(\hat{M}D_{\sigma_{k+1}}\mu'_w)(s)\\
=&\sum_{s_{k+1}\in S'}\hat{M}(s,s_{k+1})(D_{\sigma_{k+1}}\mu'_w)(s_{k+1})\\
=&\sum_{s_{k+1}\in S}M(s,s_{k+1})(D_{\sigma_{k+1}}\mu'_w)(s_{k+1})\\
=&\sum_{s_{k+1}\in S}\delta_{L(s_{k+1})\sigma_{k+1}} M(s,s_{k+1})(\mu'_w(s_{k+1}))\\
\stackrel{(b)}{=}&\sum_{s_{k+1}\in S}\delta_{L(s_{k+1})\sigma_{k+1}} M(s,s_{k+1})\sum_{L(s_0\cdots s_k)=w}  M(s_{k+1},s_k) \prod^k_{i=1}M(s_i,s_{i-1})(\mu'_0(s_0))\\
=&\sum_{ L(s_0\cdots  s_{k+1})=w \sigma_{k+1}}M(s,s_{k+1})\prod^{k+1}_{i=1}M(s_i,s_{i-1})(\mu'_0(s_0))
\end{align*}
where (b) is achieved by substituting Eq. (\ref{pp}) for $\mu'_w(s_{k+1})$. Thus we have proved Proposition 1. \qed\\

Now for $w=\sigma_0\sigma_1\cdots \sigma_k\in \Sigma^+$, the accepting probability of $\mathcal{A}$ is
\begin{align*}
&P_\mathcal{A}(w)=\Tr\left(\sum_{s\in S}\mu'_w(s)\right)\\
=&\Tr\left(\sum_{s\in S}\sum_{ L(s_0\cdots s_k)=w}  M(s,s_k) \prod^k_{i=1}M(s_i,s_{i-1})(\mu'_0(s_0))\right)\\
=&\sum_{ L(s_0\cdots s_k)=w} \Tr\left(\sum_{s\in S} M(s,s_k) \prod^k_{i=1}M(s_i,s_{i-1})(\mu'_0(s_0))\right)\\
\stackrel{(c)}{=}&\sum_{L(s_0\cdots s_k)=w}\Tr\left(\prod^k_{i=1}M(s_i,s_{i-1})(\mu'_0(s_0))\right)\\
=&P_\mathcal{M}(w)
\end{align*}
where $(c)$ holds because $\sum_{s\in S}M(s,t_k)$ is trace-preserving. This completes the proof of Lemma \ref{Lm:4}.\qed\\

Therefore, based on Theorem \ref{Th:2} and Lemma \ref{Lm:4}, we obtain the following result.
\begin{Th}The trace equivalence problem of SL-hqMC is decidable in polynomial time.
\end{Th}

\section{Quantitative analysis of linear-time properties }
Recall that $T^*$ denotes the set of all finite-length sequence over nonempty set $T$. We also need another notation $T^\omega$ that stands for the set of all infinite-length sequence over $T$.
Let $\m=(\mathcal{H}, S, M, \mu_0, AP, L)$ be an SL-hqMC. A path $\pi$ of $\m$ is an infinite sequence of states
$s_0s_1\dots\in S^\omega$ where   $M(s_{i+1}, s_{i})\neq \mathbf{0}_\h$ for all $i\geq 0$. A finite path $\widehat{\pi}$ is a finite prefix of a path.
The sets of all infinite and finite paths of $\m$ starting in state $s$ are denoted by $Path^{\m}(s)$ and $Path_{fin}^{\m}(s)$, respectively.

\subsection{Super-operator valued measure}
It is a central problem to determine the accumulated super-operator along certain paths for reasoning about the behavior  of an SL-hqMC. For example, if we can first determine the accumulated super-operator $\prod_{i=1}^k M(s_i,s_{i-1})$  in Eq. (\ref{state}), then it is easy to compute the  state $\rho_{\bar{s}}$ for an arbitrarily given initial state $\mu_0$.
To this end,  the {\it super-operator valued measure} (SVM for short) was proposed in \cite{FYY12}, which plays a similar role as probability measure for probabilistic systems. We recall the definition of SVM and some related facts as follows.

\begin{Df}
Let $(\Omega, \Sigma)$ be a measurable space; that is, $\Omega$ is a non-empty set and $\Sigma$ a $\sigma$-algebra over $\Omega$. A function
$\fdmu : \Sigma \ra \leqI$ is said to be a super-operator valued measure (SVM for short) if $\fdmu$ satisfies the following properties:
\begin{enumerate}
\item $\fdmu(\Omega) \eqsim \mathcal{I}$;
\item $\fdmu(\biguplus_i A_i) \eqsim  \sum_i\fdmu(A_i)$ for all pairwise disjoint and countable sequence $A_1$, $A_2$, $\dots$ in $\Omega$.
\end{enumerate}
We call the triple $(\Omega, \Sigma, \fdmu)$ a (super-operator valued) measure space.
\end{Df}

Given an SL-hqMC $\mathcal{M}$ and a state $s\in S$,
for any finite path
$\widehat{\pi}=s_0\dots s_n\in Path_{fin}^{\m}(s)$, we define the super-operator
$$\q(\widehat{\pi}) =
\left\{
\begin{array}{ll}
\mathcal{I},  & \mbox{if } n=0;  \\
M(s_{n}, s_{n-1})\cdots M(s_1, s_0), & \mbox{otherwise}.
\end{array}
\right.
$$
Next we define the cylinder set $Cyl(\widehat{\pi})\subseteq Path^{\m}(s)$ as
$$Cyl(\widehat{\pi})  = \{\pi\in Path^{\m}(s) : \widehat{\pi} \mbox{ is a prefix of }\pi\};$$
that is, the set of all infinite paths with prefix $\widehat{\pi}$.
Let
$$\s^\m(s) = \{Cyl(\widehat{\pi}) : \widehat{\pi}\in Path_{fin}^\m(s)\}\cup \{\emptyset\}.$$
   $Q_s$ is a mapping  from $\s^\m(s)$ to $\leqI$, defined by letting $Q_s(\emptyset)=\mathbf{0}_\h$ and
\begin{equation*}
Q_s(Cyl(\widehat{\pi}))=\q(\widehat{\pi}).
\end{equation*}

Then we have the following result.
\begin{Lm}[\cite{FYY12}]\label{thm:mainthm} The mapping $Q_s$ defined above can be extended to a SVM, denoted by $Q_s$ again, on the $\sigma$-algebra generated by
$\s^\m(s)$. Furthermore, this extension is unique up to the equivalence relation $\eqsim$.
\end{Lm}

\subsection{Linear-time Properties}

A linear-time (LT) property  over the atomic proposition set $AP$ is defined to be a subset $P$ of
$(2^{AP})^{\omega}$.
In the remainder of this section, we consider some special classes of linear-time properties. Safety is one of the most important kinds of linear-time properties.
A safety property specifies that \textquotedblleft something bad never
happens".

\begin{Df}An LT property $P$ over $AP$ is called a safety property if for all words
$\sigma\in (2^{AP})^\omega \setminus P$ there exists a finite prefix $\widehat{\sigma}\in {(2^{AP})}^*$ of $\sigma$ such that
$$P\cap\{\sigma'\in (2^{AP})^{\omega}: \widehat{\sigma} ~\text{is a finite prefix of}~ \sigma'\}=\varnothing.$$\end{Df}
Any such finite word $\widehat{\sigma}$ is called a {\it bad prefix} of $P$.
 We write $BPref(P)$ for the set of  bad prefixes
of $P$. Note that  $BPref(P)$ is a language over $\Sigma=2^{AP}$.
A safety property $P$  is called a {\it regular safety
property}, if its bad prefix set $BPref(P)$ is a regular language.

 As we know, for each regular language there exists an NFA accepting it. An NFA is a tuple $\mathcal{A}=(Q, \Sigma, \delta, Q_0, F)$ where $Q$ is finite state set, $\Sigma$ is a finite alphabet,  $\delta: Q\times \Sigma\rightarrow 2^Q$ is a transition function,      $Q_0\subseteq Q$ denotes a set of initial states, and $F\subseteq Q$ is called the accepting set. We often write $q\stackrel{a}{\rightarrow}p$ if $p\in \delta(q, a)$ where $q,p\in Q$ and $a \in \Sigma$. A string $w=a_1a_2\cdots a_n\in \Sigma^*$ is said to be {\it accepted } by $\mathcal{A}$, if there exists a finite state sequence $q_0q_1\cdots q_n$ such that $q_0\in Q_0$, $q_{i-1}\stackrel{a_i}{\rightarrow}q_i$ for $1 \leq i \leq n$, and $q_n\in F$.  The language accepted by $\mathcal{A}$, denoted by $\mathcal{L}(\mathcal{A})$, is the set of strings over $\Sigma$ that are accepted by $\mathcal{A}$.

  A DFA is a special NFA $\mathcal{A}=(Q, \Sigma, \delta, Q_0, F)$ where $|Q_0|=1$ and $|\delta(q, a)|\leq 1$ for all $q\in Q$ and $a\in \Sigma$. Then a DFA is usually denoted by $\mathcal{A}=(Q, \Sigma, \delta, q_0, F)$ where $q_0$ is the unique initial state. If it is further required that $|\delta(q, a)|= 1$ for all $q\in Q$ and $a\in \Sigma$, then $A$ is called a {\it total } DFA. Note that NFA, DFA and total DFA  accept the same language class, i.e., regular languages. In the sequel, when mentioning a DFA, we always assume it is total.

 Therefore, a safety property can be characterized by a DFA.

\subsection{Quantitative analysis of regular safety properties}
\begin{figure}[htbp]\centering
\includegraphics[width=60mm]{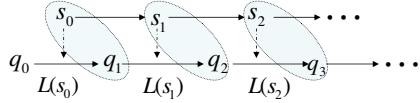}
\caption{ States in $\mathcal{M}\otimes \mathcal{A}$. }\label{Product}
\end{figure}

Given an SL-qhMC $\m=(\mathcal{H}, S, M, \mu_0, AP, L)$, a state $s\in S$, and an LT property $P$,
let $$Q_s(s\models P)=Q_s\{\pi\in Path^\m(s): L(\pi)\in P\}$$
where  $L(\pi)=L(s_0)L(s_1)\cdots$ is called the {\it trace} of  path $\pi=s_0s_1\cdots$.
In the following we show how to determine this quantity if $P$ is a regular safety property. First we note that $Q_s(s\models P)+Q_s(s\not\models P)\eqsim\mathcal{I}$ where
\begin{align*}Q_s(s\not\models P)&=Q_s\{\pi\in Path^\m(s): L(\pi)\not\in P\}\\
&=Q_s\{\pi\in Path^\m(s): pref(L(\pi))\cap \mathcal{L}(\mathcal{A})\neq \varnothing\}
\end{align*}
where $pref(A_0A_1\cdots)$ denotes the set of all finite prefixes of $A_0A_1\cdots\in (2^{AP})^\omega$, and $\mathcal{A}$ is a DFA accepting $BPref(P)$. In order to get the quantity $Q_s(s\not\models P)$, we need the concept of product between SL-hqMC and DFA.
\begin{Df}  Let $\mathcal{M}=(\mathcal{H}, S, M, \mu_0, AP, L)$ be an SL-hqMC and $\mathcal{A}=(Q, 2^{AP}, \delta, q_0, F)$ be a DFA. Then product $\mathcal{M}\otimes \mathcal{A}$ is the SL-hqMC:
$$\mathcal{M}\otimes \mathcal{A}=(\mathcal{H}, S\times Q, M', \mu'_0, \{accept\}, L')$$
where $L'(\langle s, q\rangle)= \{accept\}$ if $q\in F$ and $L'(\langle s, q\rangle)= \varnothing$ otherwise, and
\begin{align*}
\mu'_0(\langle s, q\rangle)=\left\{
         \begin{array}{ll}
           \mu_0(s), & \hbox{if $q=\delta(q_0,L(s))$;} \\
           \mathbf{0}, & \hbox{otherwise.}
         \end{array}
       \right.
\end{align*}
The transition mapping in $\mathcal{M}\otimes \mathcal{A}$  is given by
\begin{align*}
M'(\langle s', q'\rangle,\langle s, q\rangle)=\left\{
         \begin{array}{ll}
           M(s',s), & \hbox{if $q'=\delta(q,L(s'))$;} \\
           \mathbf{0}, & \hbox{otherwise.}
         \end{array}
       \right.
\end{align*}\end{Df}
For each path $\pi=s_0s_1s_2\cdots$ in $\mathcal{M}$, there exists a unique sequence of states $q_0q_1q_2\cdots$ in $\mathcal{A}$ for $L(\pi)=L(s_0)L(s_1)L(s_2)\cdots$  such that
$$q_0\stackrel{L(s_0)}{\longrightarrow}q_1\stackrel{L(s_1)}{\longrightarrow}q_2\stackrel{L(s_2)}{\longrightarrow}\cdots$$
and
$$\pi^+=\langle s_0, q_1\rangle\langle s_1, q_2\rangle\langle s_2, q_3\rangle\cdots$$
is a path in $\mathcal{M}\otimes \mathcal{A}$. Vice versa, every path in $\mathcal{M}\otimes \mathcal{A}$ which starts in state $\langle s, \delta(q_0,L(s))\rangle$ arises from the combination of a path in $\mathcal{M}$ and a corresponding state sequence in $\mathcal{A}$. The corresponding relation between the states are depicted in Fig. \ref{Product}.

Note that the DFA $\mathcal{A}$ does not affect the accumulated super-operator  along a path. That is, for each measurable set $\Pi\subseteq Path^{\mathcal{M}}(s)$,
$$Q^{\mathcal{M}}_s(\Pi)=Q^{\mathcal{M}\otimes \mathcal{A}}_{\langle s, \delta(q_0,L(s))\rangle}\{\underbrace{\pi^+: \pi\in\Pi}_{\Pi^+}\}$$
where the superscripts $\mathcal{M}$ and $\mathcal{M}\otimes \mathcal{A}$ are used to indicate the underlying systems. In particular, if $\Pi$ is the set of paths that start in $s$ and refute regular safety property $P$, i.e.,
$$\Pi=\{\pi\in Path^{\mathcal{M}}(s): pref(L(\pi))\cap \mathcal{L}(\mathcal{A})\neq\varnothing \}$$
where $\mathcal{A}$ is a DFA accepting $BPref(P)$,
then $\Pi^+$ is the set of paths in $\mathcal{M}\otimes \mathcal{A}$  that start in $\langle s, \delta(q_0,L(s))\rangle$ and eventually reach an accept state of $\mathcal{A}$:
$$\Pi^+=\{\pi^+\in Path^{\mathcal{M}\otimes \mathcal{A}}(\langle s, \delta(q_0,L(s))\rangle): L'(\pi^+)\in \lozenge accept \}.$$
Here  the linear temporal logic notation ``$\lozenge accept$'' is used to denote the LT property over $AP=\{accept\}$ consisting of  sequence $A_0A_1\cdots\in (2^{AP})^\omega$  for which there exists a finite index $i$ such that $A_i=\{accept\}$.
This shows that $Q_s(s\not\models P)$ can be derived from the accumulated super-operator for $\mathcal{M}\otimes \mathcal{A}$ reaching the set $B=S\times F$ starting from $\langle s, \delta(q_0,L(s))\rangle$. The latter problem, known as the {\it reachability problem}, can be solved by using Theorem 2.5 in \cite{FYY12}.
This is formally stated as follows.
\begin{Th} Let $P$ be a regular safety property, $\mathcal{A}$ a DFA for the set of bad prefixes $P$, $\mathcal{M}$ an SL-hqMC, and $s$  a state of $\mathcal{M}$. Then:
\begin{align*}Q_s(s\models P)&\eqsim\mathcal{I}-Q_s(s\not\models P)\\
&\eqsim \mathcal{I}- Q^{\mathcal{M}\otimes \mathcal{A}}_{\langle s, q_s\rangle} (\langle s, q_s\rangle\models\lozenge accept) \end{align*}where $q_s=\delta(q_0, L(s))$.
\end{Th}

\begin{Rm}Note that in the above procedure, we used an assumption that $\epsilon\not\in {\cal L}(A)$. Otherwise, the proof would fail. However, this is not a severe restriction since
if $\epsilon\in {\cal L}(A)$, then all finite words over $2^{AP}$ are bad prefixes, and hence, $P=\emptyset$. In this case, $Q_s(s\models P)=\mathbf{\mathbf{0}}$.
\end{Rm}

\subsection{Questions on quantitative analysis of $\omega$-regular properties}
In the above, we have taken a quantitative analysis of regular safety properties. Then it is natural to consider  the quantitative analysis of  more general LT properties; for example, how about  $\omega$-regular properties?  $\omega$-regular properties are a much larger family of LT properties than regular safety properties, and they  are characterized by B\"uchi automata.

A {\it nondeterministic B\"uchi automaton }(NBA) is represented by the same tuple $\mathcal{A}=(Q, \Sigma, \delta, Q_0, F)$ as an NFA, but with a different accepting condition. An infinite string $w=a_1a_2\cdots \in \Sigma^\omega$ is said to be accepted by NBA $\mathcal{A}$, if there exists a sequence $q_0q_1\cdots \in Q^\omega$ such that $q_0\in Q_0$, $q_{0}\stackrel{a_1}{\rightarrow}q_1\stackrel{a_2}{\rightarrow}q_2\cdots$, and $q_i\in F$ for infinitely many indexes $i\geq0$. The language accepted by NBA $\mathcal{A}$, denoted by, $\mathcal{L}_\omega(\mathcal{A})$, is the set of all infinite strings over $\Sigma$ that are accepted by $\mathcal{A}$. The class of languages accepted by NBA are called {\it $\omega$-regular languages}.
 An NBA is called  a {\it deterministic B\"uchi automaton} (DBA) if its underlying automaton is a DFA. Note that DBA can only accept a proper subset of $\omega$-regular languages.

An LT property $P$ over $AP$ is called a {\it $\omega$-regular} property, if $P$ is a $\omega$-regular language over the alphabet $\Sigma=2^{AP}$. If $P$ is a $\omega$-regular property, then its complement $(2^{AP})^\omega\setminus P$ is also a $\omega$-regular property.
Note that the regular safety property discussed before is a special case of $\omega$-regular properties. The reader can refer to \cite{BK08} for the details about $\omega$-regular properties.

Now we consider the problem of quantitative analysis of $\omega$-regular properties.
 Formally, given a state $s$ of SL-hqMC $\mathcal{M}$ and a $\omega$-regular property $P$, how to determine the following value:
$$Q_s(s\models P)=Q_s\{\pi\in Path^\m(s): L(\pi)\in P\}.$$
Here we consider a relatively simple case, that is, assume  $P=\mathcal{L}_\omega(\mathcal{A})$ for a DBA $\mathcal{A}$. Then by  similar technical treatments as before, we get that
$$Q_s(s\models P)=Q^{\mathcal{M}\otimes \mathcal{A}}_{\langle s, q_s\rangle} (\langle s, q_s\rangle\models\square \lozenge accept)$$
where $q_s=\delta(q_0, L(s))$, and ``$\square \lozenge accept$''  denotes the LT property over $AP=\{accept\}$ consisting of  sequence $A_0A_1\cdots\in (2^{AP})^\omega$  for which there exist  infinite many indexes $i$ such that $A_i=\{accept\}$. Intuitively, it means that the product system $\mathcal{M}\otimes \mathcal{A}$ starts in $\langle s, q_s\rangle$ and visits the set $S\times F$ infinitely often.

At first glance, one many think this quantity can be determined  as done for probabilistic systems \cite{BK08}. It is, however, not so easy as it looks like. In the probabilistic case, $\mathcal{M}$ is a  Markov chain and the SVM measure $Q_s$ is replaced by the probability measure $P_s$. Then,
calculating $P_s(s\models P)$ is finally reduced to finding  {\it bottom strongly connected components }(SCCs that once entered cannot be left anymore) in the underlying  graph (a digraph obtained by erasing the edge labels from the transition graph) of   Markov chain $\mathcal{M}\otimes \mathcal{A}$. The latter problem  depends only on the topological structure  and has no relation with the actual transition probability, and thus can be solved by using graph-theoretical searching algorithms. However, this method  no longer takes effect  in  quantum cases, since two states connected in the underlying  graph are not necessarily  connected in  the corresponding quantum  Markov chain. In order to  explain this point more clearly, we have a look at an example below.

\begin{figure}[htbp]\centering
\includegraphics[width=80mm]{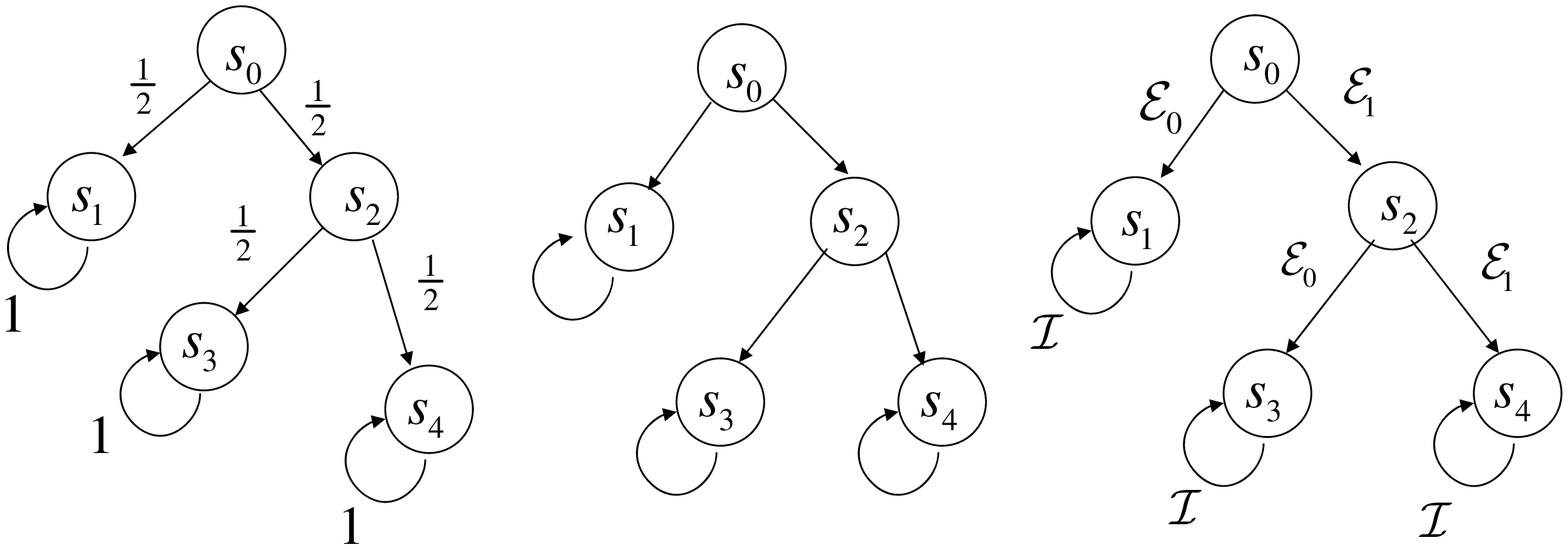}
\caption{  Difference between MC(left) and hqMC (right): Both are given by their transition graphs and the middle is their underlying graph that is obtained by erasing the edge labels.
The underlying space of hqMC is a qubit  and $\mathcal{E}_0$ and $\mathcal{E}_1$ have operation element sets $\{|0\rangle \langle 0|\}$ and $\{|1\rangle \langle 1| \}$, respectively. }\label{Graph}
\end{figure}
In Fig. \ref{Graph} the two systems have the same underlying graph.  It can be seen that  the reachability  in MC is consistent with that in its underlying graph. For example,
 $s_3$ is reachable from $s_0$ by passing $s_2$ in the  underlying graph, and at the same time $s_3$ can be reached from $s_0$ with probability $\frac{1}{4}$ in MC. However, this consistency no longer holds for hqMC. For instance,
$s_3$ is not reachable from $s_0$ in  qMC, since the accumulated super-operator along $s_0s_2s_3$ is $\mathbf{0}$.
Therefore, the quantitative analysis of general $\omega$-regular properties for quantum Markov chains cannot be addressed by using graph-theoretical algorithms. Whether this problem is solvable or not is currently not clear and needs further exploration.

\section{Conclusion}
In this paper, we have studied a novel model of quantum Markov chains. Although this model has the same expressive power as the conventional one, we have shown that it is very suited to describe hybrid systems that consist of a quantum component and a classical one.  Based on the quantum Markov chain model, we have further proposed an automaton model called hybrid quantum automata that can be used to describe  hybrid systems that receive input (or actions) from the outer world. The language equivalence problem of hybrid quantum automata has been shown to be decidable in polynomial time, and furthermore we have applied this result to the trace equivalence problem of quantum Markov chains which is thus also decidable in polynomial time. Finally, we have discussed model checking linear-time properties of hybrid systems, showing that the quantitative analysis of regular safety properties can be addressed successfully as done for stochastic systems, but  the problem for  general $\omega$-regular properties is more difficult and needs further exploration.

Hybrid systems modeled by quantum Markov chains have already been often encountered in quantum information processing, and the quantum engineering systems developed in the  future will most probably be hybrid systems.  Therefore, it is worth developing a theory for describing and verifying these hybrid systems. We hope this work can stimulate further discussion.

\section*{Acknowledgement} The authors are thankful to the  anonymous referee
for valuable comments and suggestions, and to Professor Mingsheng Ying for carefully reading the manuscript and giving some helpful suggestions. Especially, the first author is greatly indebted to Professor Ying for his guidance when the first author was working in QCIS of UTS.
  This work is supported in part
by the National Natural Science Foundation of China (Nos. 61100001, 61472452, 61272058, 61428208 and 61472412), the  National Natural Science Foundation of Guangdong province of China (No. 2014A030313157), Australian Research Council under Grant DP130102764, and the CAS/SAFEA International Partnership Program for Creative Research Team.

\end{document}